\documentclass[aps,prb,twocolumn,showpacs]{revtex4}

\usepackage{graphicx}
\usepackage{dcolumn}
\usepackage{bm}

\begin{document}


\title{Classical heisenberg antiferromagnet away from the pyrochlore lattice limit:\\
entropic versus energetic selection\\
}


\author{C. Pinettes}
\affiliation{Laboratoire de Physique Th\'eorique et Mod\'elisation, Universit\'e de Cergy-Pontoise,\\
5 Mail Gay-Lussac, Neuville-sur-Oise, 95031 Cergy-Pontoise Cedex, France\\
and 2. Physikalisches Institut, RWTH-Aachen, 52056 Aachen, Germany}

\author{B. Canals, C. Lacroix}
\affiliation{Laboratoire de Magn\'etisme Louis N\'eel
CNRS, B.P. 166, 38042 Grenoble Cedex 9, France}

\date{\today}

\begin{abstract}
The stability of the disordered ground state of the classical Heisenberg 
pyrochlore antiferromagnet is studied within extensive Monte Carlo simulations
by introducing an additional exchange interaction $J'$ that interpolates between 
the pyrochlore lattice ($J'=0$) and the face-centered cubic lattice ($J'=J$).
It is found that for $J'/J$ as low as $J'/J\ge 0.01$, the system is long range
ordered : the disordered ground state of the pyrochlore antiferromagnet
is unstable when introducing very small deviations from the pure $J'=0$ limit.
Furthermore, it is found that the selected phase is a collinear state 
energetically greater than the incommensurate phase suggested by
a mean field analysis.
To our knowledge this is the first example where entropic selection prevails over
the energetic one.
\end{abstract}


\maketitle

\section{\label{introduction}Introduction}

In recent years, there has been a great interest in the study of 
fully frustrated Heisenberg antiferromagnets (HAF),
both from the experimental and theoretical point of view\cite{Ramirez,Diep94}.
These systems often show unconventional ground-states, noncollinear,
incommensurate\cite{Ferey86} or disordered\cite{Booth00}.
In these systems, the classical Mean Field (MF) description indicates a 
macroscopic ground-state degeneracy and no long range order is predicted.
This property is directly related to the extensive number of degrees of freedom
that still fluctuate at low temperature\cite{Moessner98}.
Nevertheless, in many cases, thermal or quantum fluctuations give rise to an 
entropic selection within the ground state manifold, known 
as ``order by disorder''\cite{Villain80}.
This phenomenon may induce a selection of a subset of the continuous degenerate
ground states as it does for the kagom\'e antiferromagnet\cite{Chalker92}, 
and sometimes can even select a particular ground
state as it is the case for the face centered cubic (fcc) 
antiferromagnet\cite{Henley87}.

\begin{figure}
\includegraphics[width=7cm]{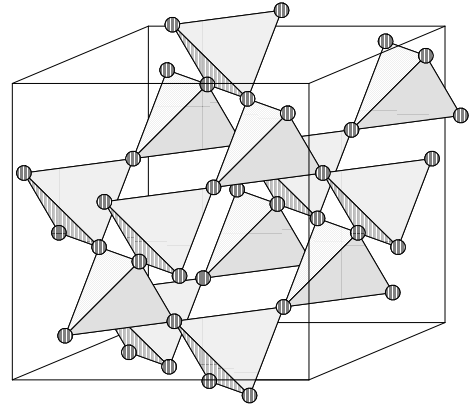}%
\vspace{0cm}
\caption{\label{fig1}The pyrochlore lattice.}
\end{figure}

Among these systems, the 3D pyrochlore lattice is of particular interest 
since it does not display such ``order by disorder'' and remains disordered
down to zero temperature\cite{Villain80,Reimers92,Moessner98}.
It consists in a three dimensional arrangement of corner-sharing tetrahedra
(see Fig.~\ref{fig1}). 
All magnetic compounds which cristallize in the pyrochlore structure 
exhibit unusual magnetic properties. 
A few order at well defined N\'eel temperature \cite{Ferey86} 
while many of them behave as spin glasses despite the high 
degree of stoichiometry\cite{Booth00}.
In the Heisenberg antiferromagnetic $S=1/2$ quantum case, 
the pyrochlore lattice exhibits no magnetic transition even at T=0K,
and behaves as a spin-liquid.
The spin-spin correlation length never exceeds the inter atomic distance and
it is expected that there is a continuum of singlet states within 
the singlet-triplet gap\cite{Canals98}.
For classical Heisenberg spins, Monte Carlo simulations have shown that 
the pyrochlore HAF also behaves as a spin-liquid, with short ranged
spin correlations both in space and time\cite{Reimers92,Moessner98},
which allows for its description to take into account only
the tetrahedral unit cell\cite{Moessner99}.
The exactly solvable infinite-component spin vector model on the pyrochlore 
lattice also does not order down to zero temperature\cite{Canals01}.
Classical models providing much informations close to the characteristics 
obtained in the quantum case, it indicates that the low-energy physics in 
the pyrochlore structure is mainly related to the geometry of the lattice.

An interesting aspect of the problem is to study how this disordered 
classical ground-state resists to perturbations as it could enlarge
its observation in real compounds. 
It has been shown that ordering is very easily induced 
either by including further neighbour interaction\cite{Reimers91,Harris92}
or easy plane anisotropy\cite{Bramwell94,Moessner98} 
or lattice deformations\cite{Terao96}.
The purpose of this paper is to explore the effect of coupling 
the spins of the pyrochlore net to the remaining nearest-neighbor (n.n) 
spins in the fcc net via standard classical Monte Carlo (MC) calculations.
The question is wether the system will select an ordered state when 
switching on this coupling or maintain disorder as in the pyrochlore
HAF.
As obtained in previous works, an energetic comparison of all magnetic phases
at the MF level already lifts the degeneracy and strongly supports that 
any deviations from the pyrochlore limit will drive the system to an ordered state 
(sec.~\ref{pyro-fcc}).
But strikingly, it is shown that thermal fluctuations do not select the expected
mean field ground state but prefer a collinear phase which is more stable from
the entropic point of view (sec.~\ref{mc-results}).
\section{\label{pyro-fcc}From the pyrochlore lattice to the fcc lattice : the $J-J'$ model}
\begin{figure}
\includegraphics[width=7cm]{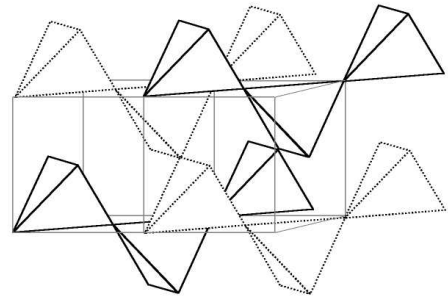}%
\vspace{-0.5cm}
\caption{\label{fig2}The $J-J'$ n.n. fcc lattice.
The fcc lattice is decomposed into two pyrochlore nets :
$A$ sites (plain lines) and $B$ sites (dashed lines). 
$J$ couples $A-A$ n.n. sites (plain lines) 
while $J'$ couples $B-B$ n.n. sites (dashed lines) 
and $A-B$ n.n. sites (not drawn for clarity).
This model contains 8 sites per unit cell.}
\end{figure}

In both fcc and pyrochlore lattices, the classical energy is minimized by any configuration
for which the total spin of each elementary tetrahedron in the lattice is zero. 
The main
difference in the ground-states of these two lattices comes from the fact that in the fcc 
net the ordering at one tetrahedron will select a unique global ordering, 
while in the pyrochlore net, since the tetrahedra are more sparsely connected,
there remains an extensive number of degrees of freedom\cite{Moessner98}.
In the fcc HAF, the MF ground-states are degenerate
along the $\pi/a(1,q,0)$ direction in reciprocal space.
Thermal fluctuations break this degeneracy 
and select the collinear ordered phase ${\bf q}_{0}=(\pi/a,0,0)$\cite{Henley87}. 
These results have been confirmed by standard Monte Carlo simulations\cite{Minor88}
which show evidence for a first order phase transition.
In the pure pyrochlore HAF, the MF ground-states
are degenerate throughout the entire Brillouin zone.
Numerical works have shown that no long-range order occurs at nonzero 
termperature in that case, 
even in a regime where the temperature is below the energy scale 
set by interaction\cite{Reimers92,Moessner98}.
In between these two limits, we have considered the $J-J'$ fcc HAF, 
where $J$ couples the n.n. spins on a pyrochlore net and 
$J'$ the remainder n.n. spins on the underlying fcc net (see Fig.~\ref{fig2}).
This leads to the following classical Heisenberg hamiltonian :
\begin{equation}
H = - J \sum_{<ij>}  {\bf S}_{i}.{\bf S}_{j} - J' \sum_{<i'j'>}  {\bf S}_{i'}.{\bf S}_{j'} 
- J' \sum_{<ii'>}  {\bf S}_{i}.{\bf S}_{i'}
\end{equation}
where ${\bf S}_{i}$ is a vector spin of unit length occupying the $i^{\rm th}$ fcc lattice 
site, $i$ and $i'$ design sites on two pyrochlore nets 
shifted by $(2a,0,0)$ respectively and  the sums $< >$ run over n.n. pairs.
We have limited our study to the antiferromagnetic case, i.e. all negative interactions.
$J=J'<0$ corresponds to the fcc HAF and $J'=0$, $J<0$ leads to the pyrochlore HAF.
The cubic lattice constant is $2a$. 

We have first studied the MF ground-states degeneracy for small $\alpha=J'/J$. 
One way to do so is to compute the Fourier transform of the interactions 
on the lattice, $J_{\bf q}$.
The ground-state (if unique) then corresponds to the highest eigenvalue 
$\lambda_{max}({\bf q}_{1})$  of $J_{\bf q}$ and is described by 
the propagation vector ${\bf q}_{1}$\cite{Bertaut63}.
For the $J-J'$ model, $J_{\bf q}$ is a $8\times8$ matrix that has already been calculalted
by Reimers {\it et al.}\cite{Reimers91}.

\begin{figure}
\includegraphics[width=7.5cm]{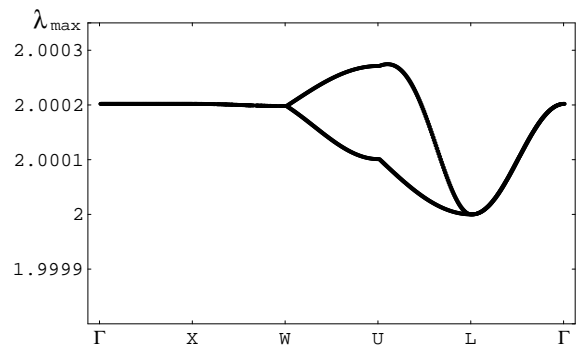}%
\caption{\label{fig3}Largest eigenvalue $\lambda_{max}({\bf q})/|J|$  along the high 
symmetry axis of the fcc Brillouin zone for $\alpha=0.01<\alpha_{c}\approx 0.21$.}
\end{figure}

For $0<\alpha<\alpha_{c}\approx 0.21$, we have reported the largest eigenvalue 
$\lambda_{max}({\bf q})/|J|$ along the high 
symmetry axis of the fcc Brillouin zone on figure~\ref{fig3}.
The degeneracy is completely lifted and $\lambda_{max}({\bf q})$ 
has an absolute maximum ${\bf q}_{1}$ along the $(0,q,q)$ direction, 
selecting an incommensurate phase
of wave vector ${\bf q}_{1}$. 
Within the MF approximation the $J-J'$ model selects an ordered phase for $0<\alpha<\alpha_{c}$.

For $\alpha\ge \alpha_{c}$, the situation is very similar to the fcc case. 
The MF degeneracy is lifted except on lines along $\pi/a(1,q,0)$ direction : the system
remains disordered in that case. 

\begin{figure}
\includegraphics[width=7.5cm]{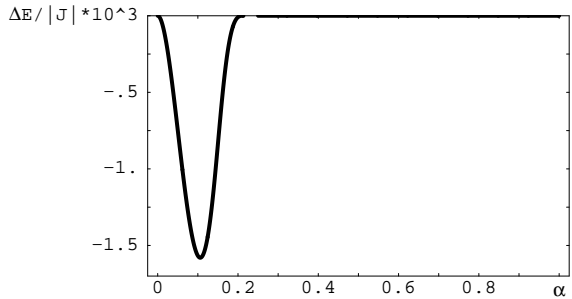}%
\caption{\label{fig4}MF energy difference $10^3\times\Delta E/|J|$ between the two MF ground-states
as a funtion of $\alpha$ ($\Delta E=E({\bf q}_{1})-E({\bf q}_{0})$, see text).}
\end{figure}

We have reported in figure~\ref{fig4} the MF energy difference 
$\Delta E/|J|$ between the two ground states:
the ${\bf q}_{1}$ incommensurate state and the $\pi/a(1,q,0)$ states 
(including the ${\bf q}_{0}$ collinear state) as a funtion of $\alpha$.
We see that $\Delta E/|J$ is much lower than the energy scale set by the interaction 
($\Delta E/|J|\sim \alpha/100$). 
Thus the degeneracy is very slighty lifted for $\alpha<\alpha_{c}$ : 
the highest branch is almost flat over the whole Brillouin zone 
and very similar to the pyrochlore one.
If we add thermal fluctuations 
on these MF ground-states, it is not clear what will happen. 
They can either select a particular state, 
the ${\bf q}_{1}$ noncollinear state (energetic selection) 
or a collinear state (entropic selection) or induce disorder. 
\section{\label{mc-results}Monte Carlo results}
We have performed standard MC simulations using the sample sizes of $N=864-6912$ spins 
with periodic boundary conditions, over a temperature range extending down to $T/|J|=10^{-4}$. 
We started from both random and ordered initial configurations. 
In each run, up to $10^7$ MC steps per spin have been discarded for equilibrating 
and up to $2\times10^7$ MC steps per spins have been used for averaging.
We have checked that our simulations reproduce the Monte-Carlo results 
for the pyrochlore net\cite{Moessner98} and those for the fcc net\cite{Minor88}.
In order to determine unambigously whether the $J-J'$ fcc HAF order or not, 
we have measured more specifically the specific heat per spin and a correlation function 
which quantify the collinearity of spins.

An unbiaised way to look for fluctuations induced order is to measure heat capacity,
 since it is very sensitive to the presence of zero modes (modes whose energy is 
independent of displacement to second order at all wave vectors)\cite{Chalker92,Moessner98}.
From the classical equipartition theory, quadratic modes 
contribute a factor $T^{1/2}$ to the partition function Z, while quartic modes 
contibute a factor $T^{1/4}$ : at low temperature,
the behavior is thus dominated by fluctuations around the ground-states with the largest 
number of zero modes. These zero modes leave a signature in the heat capacity
(since they contribute to $k_{B}/4$ instead of $k_{B}/2$ for the quadratic modes)
that can be checked with Monte Carlo simulations. 

In the n.n. pyrochlore HAF, fluctuations around the collinear states
lead to one zero mode per tetrahedron 
resulting in a heat capacity per spin $C=3k_{B}/4$ in absence of order
(instead of $C=5k_{B}/8$ if a collinear state is selected). This value has already been
checked by Monte Carlo calculations\cite{Reimers92,Moessner98}.
In the fcc n.n. HAF, fluctuations select the collinear state  and 
since it contains only degeneration lines, the heat capacity per spin is simply equal to 
$C=k_{B}$ in that case.

\begin{figure}[h]
\includegraphics[width=8cm]{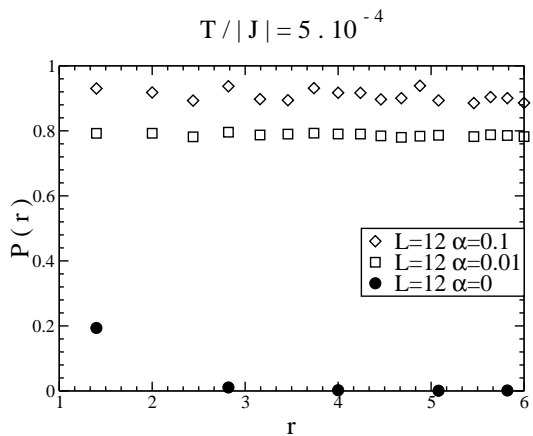}%
\caption{\label{fig5}Dependence of the collinearity parameter $P(r)$ at $T/|J|=5\times10^{-4}$
for different values of $\alpha=J'/J$, $4\times L^{3}$, $L=12$ Heisenberg spins
and starting from random spin configurations. 
$r$ is given in units of n.n. distances.}
\end{figure}

\begin{figure}
\includegraphics[width=8cm]{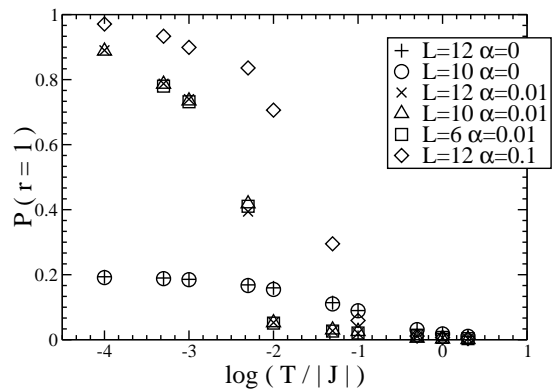}%
\caption{\label{fig6}Temperature dependence of the collinearity parameter $P(r=1)$ 
for different values of $\alpha=J'/J$, different sizes $4\times L^{3}$ and 
starting from random spin configurations.}
\end{figure}

\begin{figure}
\includegraphics[width=8cm]{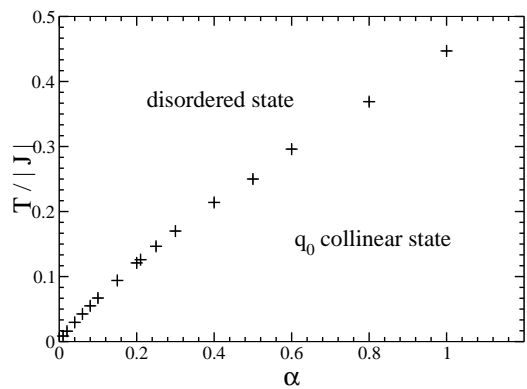}%
\caption{\label{fig7}Phase diagram of the $J-J'$ model versus  $\alpha=J'/J$ (${\bf q}_{0}=(\pi/a,0,0))$.}
\end{figure}

As the heat capacity derived from energy fluctuations by MC simulations is rather noisy 
at low temperatures, we have estimated the heat capacity directly from 
the slope of the asymptotic internal energy at low temperatures. 
\begin{table}%
\caption{\label{chaleur-specifique}Extrapolated zero temperature specific heat for various values of $\alpha=J'/J$ from $\alpha=0$ (pyrochlore HAF) to $\alpha=1$ (fcc HAF).
This system displays long range order($C~=~1$) for all $\alpha > 0$.}
\begin{ruledtabular}
\begin{tabular}{llll}
 & $\alpha=J'/J$ 	& $C \rm{~in~} k_B \rm{~units}	$ & \\ \hline
 & \vspace{-0.2cm}	&				& \\
 & 0 (pyrochlore HAF)	& 0.73$\pm$0.03			& \\
 & 0.01			& 1.01$\pm$0.02			& \\
 & 0.1			& 1.00$\pm$0.02			& \\
 & 1 (fcc HAF)		& 0.998$\pm$0.005		& \\
\end{tabular}
\end{ruledtabular}
\end{table}
The obtained results are reported in Table~\ref{chaleur-specifique} and
clearly show that for $\alpha\ge 0.01$, thermal fluctuations select an 
ordered state at zero temperature.
To show more directly whether the system selects the collinear ordering or not, 
we have calculated a parameter that measures the collinearity\cite{Moessner98} :
\begin{equation}
P(r_{i,j})=\frac{3}{2} \left( \begin{array}{c}
<({\bf S}_{i}.{\bf S}_{j})^2> - \frac{1}{3}\\
\end{array} \right)
\end{equation}
where $r_{i,j}$ is the distance between $i$ and $j$ sites.
So P=1 for collinear spins and P=0 in the high temperature limit.
This parameter is represented in figure \ref{fig5} as a function of the separation r 
(in units of n.n. distances) at $T/|J|=5\times10^{-4}$ for different values of $\alpha$.
We find that as soon as $\alpha\ge0.01$, the collinearity parameter is of long range,
and close to 1, contrary to the pyrochlore limit which shows a correlation for $r<2$ only.
We can note that the order parameter $P(r \rightarrow \infty)$ is rather lower than 1
despite the low-temperature taken, as for the XY model on the pyrochlore 
antiferromagnet\cite{Moessner98}.
Figure~\ref{fig6} shows the temperature dependence of the collinearity parameter for neighbouring spins $r=1$.
As soon as $\alpha\ge0.01$, the spins become collinear in the low temperature limit, by contrast
with the pyrochlore case which has a small but finite correlation length.
We have reported in figure~\ref{fig7} the critical temperature as a function of $|J'|$ 
for $J'\neq0$.
It is of the order of $|J'|/2$.
We have checked that these results do not depend on the initial configuration choosen, 
by starting from collinear and random states. 
And the sizes considered are large enough since no marked finite-size effects are observed.

We have checked the low temperature phase starting from collinear and random states. 
We found the ${\bf q}_{0}$ collinear state for $\alpha\ge 0.01$.
Let's notice that starting from a random configuration, stacking faults set in 
at low temperatures in MC calculations, as it is usual in fully frustrated systems\cite{Minor88}.
At $T/|J|=5\times10^{-4}$ and for $\alpha=0.01$ and $N=6912$, we found two collinear blocks 
(of ${\bf q}_{0}$ type) of around $N/2$ spins and two non collinear walls of few hundred spins.
For $\alpha\ge \alpha_{c}$, we observe an order by disorder effect very similar to the fcc HAF : 
thermal fluctuations induces the collinear magnetic order ${\bf q}_{0}$ via an entropic selection. 
For $0<\alpha<\alpha_{c}$, the low temperature state selected by thermal fluctuations 
is not the ${\bf q}_{1}$ incommensurate state selected at the MF level, neither a
state with a MF energy close to the ${\bf q}_{1}$ state and commensurate with the
periodic boudary conditions we have taken, but the collinear state ${\bf q}_{0}$ 
whose wave vector is far from ${\bf q}_{1}$ in the Brillouin zone.
Even for $\alpha\approx 0.1$, where the MF energy difference between these two ground states 
becomes greater ($\Delta E/|J|\sim 10^{-3}$, see Fig.~\ref{fig4}) than the lowest temperature we have 
considered $T/|J|=5\times10^{-4}$, the system still selects the collinear state.
The entropic selection always overcomes the energetic one in that case.
\section{\label{conclusion}Conclusion}
We have studied in this paper the effect of small deviations from the pyrochlore n.n. HAF
to the fcc one via standard MC simulations. 
As soon as we switch on $J'$ ($\alpha=J'/J\ge0.01$), correlation functions are long-ranged 
in space and the specific heat at zero temperature does not indicate the presence of zero modes.
The $J-J'$ HAF selects the ${\bf q}_{0}=(\pi/a,0,0)$ collinear state at low-temperature 
and ordering occurs at finite temperature.
For small $\alpha$, thermal fluctuations 
do not select the MF ground-states at low temperatures but the collinear state. 
The entropic selection prevails over the energetic one. 
This is the first case, to our knowledge, where such a phenomenon is pointed out.
The pyrochlore lattice thus remains the only 3D example known with classical
Heisenberg spins remaining disordered at all temperatures.
\begin{acknowledgements}
One of us (C.P.) wishes to thanks Professor H.T. Diep for helpful discussions 
and Professor G. Guentherodt for hospitality at the 2. Physikalisches Institut, Aachen. 
This work was supported in part by the DFG through SFB341.
'Laboratoire de Physique Th\'eorique et Mod\'elisation' is associated with CNRS (ESA 8089).
\end{acknowledgements}

\end{document}